\documentstyle[12pt]{article}

\advance \topmargin by -\headheight
\advance \topmargin by -\headsep

\evensidemargin \oddsidemargin
\begin{document}

%%%%%%%%%%%%%%%%%%%%%%%%%%%%  MACROS.TEX  %%%%%%%%%%%%%%%%%%%%%%%%%%%%%%%%%%
\def\rf#1{(\ref{eq:#1})}            \def\lab#1{\label{eq:#1}}
\def\foot#1{\footnotemark\footnotetext{#1}}
\def\br{\begin{eqnarray}}           \def\er{\end{eqnarray}}
\def\be{\begin{equation}}           \def\ee{\end{equation}}
\def\({\left(}                      \def\){\right)}
\def\nonu{\nonumber}
\def\norm{\normalsize\baselineskip=26pt}
%  \def\beq{\begin{equation}}      \def\eeq{\end{equation}}
%  \def\bea{\begin{eqnarray}}      \def\eea{\end{eqnarray}}

%%%%%%%%%%%%%%%%%%%%%%%%%%%%% math symbols  %%%%%%%%%%%%%%%%%%%%%%%%%%%%%%%
\def\a{\alpha}
\def\b{\beta}                       \def\bt{\tilde \beta}
\def\d{\delta}                      \def\D{\Delta}
\def\eps{\epsilon}
\def\g{\gamma}                      \def\G{\Gamma}
\def\grad{\nabla}
\def\h{{1\over 2}}	\def\i{{\rm i}}  	\def\e{{\rm e}}
\def\k{\kappa}
\def\l{\lambda}                     \def\L{\Lambda}
\def\m{\mu}
\def\n{\nu}
\def\o{\omega}                      \def\O{\Omega}
\def\p{\phi}                        \def\P{\Phi}        \def\vp{\varphi}
\def\pa{\partial}
\def\pr{\prime}
\def\ra{\rightarrow}
\def\s{\sigma}                     % \def\S{\Sigma}
\def\t{\tau}
\def\th{\theta}   \def\Th{\Theta}   \def\vth{\vartheta}  \def\va{\vartheta}
\def\ti{\tilde}
\def\wti{\widetilde}
\def\vareps{\varepsilon}
\def\veps{\varepsilon}
\def\tg{\bigtriangleup}

\newcommand{\ct}[1]{\cite{#1}}
\newcommand{\bi}[1]{\bibitem{#1}}
\renewcommand{\theequation}{\thesection.\arabic{equation}}

%%%%%%%%%%%%%%%%%%%%%%%%%%%%%%% Title Page %%%%%%%%%%%%%%%%%%%%%%%%%%%%%%%%%
\begin{titlepage}
\vspace*{-1cm}
\begin{flushright}
UB-TH-0193 \\
hep-th/9307175 \\
Revised: March 1, 1994
\end{flushright}
\vfil
\begin{center}
{\Large\bf Multi--Soliton Solutions of Affine Toda Models}
\end{center}
\vfil
\begin{center}
{\large Z. Zhu   and   D. G. Caldi}\\
\vskip 20pt
{Department of Physics} \\
{State University of New York at Buffalo} \\
{Buffalo, N.Y. \ \ \ 14260-1500}\\
\end{center}
\vfil
\baselineskip=26pt
\begin{center}
{\large {\bf ABSTRACT}}\\
\end{center}
\vfil

\noindent
Hirota's  method is used to
construct multi--soliton and plane--wave solutions for  affine Toda
field theories  with imaginary coupling.

\vfil
\end{titlepage}

\baselineskip=26pt
%%%%%%%%%%%%%%%%%%%%%%%%%%%%% Introduction %%%%%%%%%%%%%%%%%%%%%%%%%%%%%%%%%
\section{Introduction}
\setcounter{equation}{0}
There has been a revived interest in Toda field theories [1-12]
recently, chiefly because they represent a large class of integrable
two-dimensional models which include not only conformal field theories
(CFT's) but also massive deformations away from conformality.  Toda
field theories (TFT's) have a Lagrangian realization (and hence
provide one for the CFT's which can be written as TFT's) and can be
classified according to their underlying Kac-Moody algebra $g$. If $g$
is a finite-dimensional, semi-simple Lie algebra, the theory is a CFT.
If $g$ is an affine Kac-Moody loop algebra, the resulting affine Toda
field theory is massive, but still integrable. Indeed, these theories
provide explicit realizations of the phenomenon investigated by
Zamolodchikov \ct{zam} concerning integrable massive deformations of
CFT's.  Finally, it has been observed \ct{BB,ACFGZ} that if one has
the full affine Kac-Moody algebra (i.e., with central extension and,
in fact, two more fields compared with the usual TFT), then the theory
is conformal again and is called a conformal affine Toda field theory.
The affine TFT is then considered a ``gauge-fixed'' version of the
conformal affine theory.

In this paper we are concerned with finding exact solutions to
classical affine TFT's.  To this end we use the method introduced by
Hirota \ct{hir} and applied to TFT's in the stimulating work of
Hollowood \ct{hollo1}.  If the coupling in the theory is real, then
the Toda fields can be consistently taken as real fields. The theories
are unitary and the S-matrices have been solved \ct{bcds}.  Since the
interaction potential has a unique minimum, there are no solitons.
But these real affine TFT's are not those related to unitary CFT's.
Curiously enough, it turns out that to have the relationship to CFT's
mentioned above, one must take the coupling constant to be purely
imaginary \ct{hollo1}.  But then one has the additional feature that
the Toda potential is periodic, i.e., invariant under translations on
the weight lattice of $g$.  Hence, the vacua are infinitely degenerate
and there is a rich spectrum of topological solitons. But the
Lagrangian is complex and the Hamiltonian is non-Hermitian; so the
theory would appear to be non-unitary. Yet the solitons have real
masses, and it has been suggested that for some region of the
imaginary coupling, there may exist a truncation which would render
the theories unitary \ct{hollo2}.

Hirota's method has been successfully applied to the affine Toda
theories with purely imaginary coupling.  Single--soliton solutions
were obtained in \ct{hollo1,mm,ACFGZ} and multi--soliton solutions for
$A_r^{(1)}$ theories in \ct {hollo1}.  In this paper, we will
construct more multi-soliton solutions to the complex affine Toda
theories.

In \S 2, we will review briefly the affine Toda theories and Hirota's
method.  Since the solutions for non--simply--laced Toda models can be
derived from the simply--laced ones, we will mostly work on the
simply--laced cases with some comments on non--simply--laced cases.
In \S 3, we construct multi--soliton solutions for the $\hat A_r$ loop
algebras, in \S 4, we deal with $\widehat so$(8), and in \S 5, $\hat
D_{2r}$, We summarize our results and conclusions in \S 6.

%%%%%%%%%%%%%%%%%%%%%%%%%%%%% Method %%%%%%%%%%%%%%%%%%%%%%%%%%%%%%%%%%%%%%%%%
\section {Affine Toda Models and Hirota's Method}
\bigbreak
The Lagrangian density of affine Toda field theory can be written in the form
\be
  {\cal L}={1\over2}(\partial_\mu \vp)(\partial^\mu\vp) -
  {2\over\psi^2}{m^2\over\bt ^2}\sum_{j=0}^r n_j(\e^{\bt\a_j\cdot\vp}-1).
  \lab{lag}
\ee
The field $\vp(x,t)$ is an $r$-dimensional vector, $r$ being the rank
of some finite dimensional semi--simple Lie algebra $g$.  The
$\alpha_j$'s, for $j=1,...,n$ are the simple roots of $g$; $\psi$ is
the highest root: $\psi=\sum_{j=1}^r n_j\a_j$, where the $n_j$'s are
positive integers, and $n_0=1$.  $\a_0=-\psi$ is the extended root of
$\hat g$, the affine extension of $g$.  $\bt$ is the coupling
constant, m is the mass parameter. Note, in general repeated indices
are not summed over in this paper.

The equations of motion are
\be
  \partial^2 \vp = - {2\over \psi^2} {m^2\over\bt}\sum_{j=0}^r n_j \alpha_j
  \e^{\bt \alpha_j\cdot\vp}.   \lab{eom1}
\ee
Writing $\vp$ out  in  component form,
$\quad \vp =  \sum_{j=1}^r({2 \a_j/\a_j^2})\, \vp^j $,
we have
\be
   \pa^2 \,\vp^j =-{m^2\over \bt} l_j \( \exp ({\bt \sum_{k=0}^r K_{jk}\vp^k})
       -\exp ({\bt \sum_{k=0}^r K_{0k} \vp^k }) \), \lab{eom2}
\ee
where $K_{jk}=2 \a_j \cdot \a_k /\a_k^2$ is the the Cartan matrix of $\hat g$,
and $l_j$'s are integers related to $n_j$'s by
$ l_j = ({\a_j^2/ \psi^2})\, n_j$,  \  for  j = 0, 1, $\cdots$, r.
As noted in the introduction, we are interested in the case where  the
coupling constant is purely imaginary,
        $$ \bt = \i \b. $$

We now use Hirota's method to solve the equations of motion.
One introduces a nonlinear transformation
\be
   \vp = -{1\over \i\b}\sum_{j=0}^r {2\a_j\over \a_j^2} \ln\tau_j,
     \quad \mbox{or} \quad
   \vp^j = - {1\over \i\b}\Bigl( \ln{\t_j - l_j \ln \t_0} \Bigr), \lab{transf}
\ee
where  the new variables are the tau--functions for  the theories.
The equations of motion \rf{eom1}  then  decouple to a collection
of equations in  Hirota's form,
\be
   {1\over 2 m^2}\, (D_t^2-D_x^2)\,\tau_j\cdot\tau_j =
   l_j \left ( \prod_{k=0}^r \tau_k^{-K_{jk}}-1 \right )\tau_j^2.  \lab{eom3}
\ee
Here  $D_x$ and $D_t$ are Hirota derivatives, defined by
   $$ D_x^m D_t^n\, f\cdot g=\left ({\pa \over \pa x}- {\pa
       \over \pa x^{\prime}} \right ) ^m \left ({\pa \over \pa
       t}- {\pa \over \pa t^{\prime}} \right ) ^n
       f(x,t)g(x^{\prime},t^{\prime}) \, \bigg\vert  _{x=x^{\prime}
      \atop t=t^{\prime}}. $$
We expand the $\tau$-functions in a formal power series in an arbitrary
parameter  $\epsilon$ as
\be
   \tau_j = \sum_{n=0} \epsilon^n \tau^{(n)}
          = 1 + \epsilon \tau^{(1)}_j + \epsilon^{2} \tau^{(2)}_j + \cdots.
   \lab{tauexpansion1}
\ee
Using the bilinearity of  Hirota's operator, we can rewrite
\rf{eom3}  as
\be
   {1\over 2 m^2} \sum_{n} \sum_{a+b=n} {\epsilon^n
    (D_t^2-D_x^2)\,\tau_j^{(a)}\cdot \tau_j^{(b)}} =
    l_j \left ( \prod_{k=0}^r \tau_k^{-K_{jk}}-1 \right )\tau_j^2 . \lab{eom4}
\ee
Expanding  the r.h.s.\ also as power series in $\epsilon$, and equating
corresponding coefficients, the resulting equation can be written in the
form:
\be
   \sum_k L_{jk}\t_k^{(n)} + {1\over m^2} (\pa_t^2 - \pa_x^2)\, \t_j^{(n)}
    = U_j^{(n)}, \lab{pde}
\ee
where $L_{jk} = l_j K_{jk} $, and
$U_j^{(n)}$ collects  the rest of the $n^{\rm th}$ order terms
which depend on $\t^{(1)}$, $\cdots$, $\t^{(n-1)}$.
The explicit form of $U_j^{(n)}$ depends on the Cartan matrix K.
Now the nonlinear equations  of motion are simplified to a set of
linear partial differential equations, which can be solved iteratively.
Furthermore, the first order equation is homogeneous, $U_j^{(1)} = 0$.

For a single soliton traveling at velocity v, $\t_j$ should depend on x and t
only through the combination $z=x - vt$,
    $ \t_j(x,t) = \t_j(x-vt) = \t_j(z) $.
Here equation \rf{pde} simplifies to a set of ordinary differential
equations:
\be
   \sum_k L_{jk} \t_k^{(n)} (z) -{1-v^2 \over m^2}
    {{\rm d}^2\over {\rm d}\,z^2} \t_j^{(n)}(z)
    = U_j^{(n)} ,  \lab{ode}
\ee
with the first order equation given by:
\be
   (1-v^2){{\rm d}^2\over {\rm d}\,z^2} \t_j^{(1)}(z) =
    m^2 \sum_k L_{jk} \t_k^{(1)}(z).\lab{ode1}
\ee

Equation \rf{ode1} can be solved by diagonalizing the matrix $L_{jk}$.
There are two linearly independent solutions for any given eigenvector
$\L$ (with corresponding eigenvalue $\l$).  For the null eigenvector
$\L_0 = ( l_0, l_1, \cdots, l_r) $, with eigenvalue $\l_0=0$,
the two linearly independent solutions are:
\be
  \t_j^{(1)}(z) = (c_1 + c_2 z) l_j, \lab{sol1}
\ee
which correspond to vacuum solutions.  For $\L$ with non--zero
eigenvalue $\l$, the two linearly independent solutions are:
\be
   \t_j^{(1)}(z) = \L_j \e^{\g z+\xi},  \lab{sol2}
\ee
where $\xi$ is an integration constant, and $\g$ satisfies
${(1-v^2) \g^2} = m^2 \l $.   Since $\g$ can be either positive or
negative, \rf{sol1} and \rf{sol2} give all the  linearly independent
solutions to \rf{ode1}.

General solutions to \rf{ode1} are superposition of the solutions given
in \rf{sol2}, solution to \rf{ode} can also be superposed with a solution
to \rf{ode1}, but they will not have properties of a single soliton, and
will be dealt with later as multi--soliton solutions.
Here we will restrict our solutions to the single--soliton form:
\be
    \t_j^{(n)} = \d_j^{(n)} \e^{n\G},  \lab{ansatz}
\ee
where $ \G=\g(x-vt)+\xi$, and $\d_j^{(n)}$ is independent of x and t.
Now \rf{ode} can be simplified to a set of linear algebraic equations:
\be
   \sum_k \( L_{jk} - n^2 \l \d_{jk} \) \d_k^{(n)} = V_j^{(n)}, \lab{alg}
\ee
where $V_j^{(n)}$ depends only on $\d^{(1)}$, $\cdots$, $\d^{(n-1)}$.
In particular, $V_j^{(1)} = 0$, so $\d^{(1)}$ is an eigenvector of L
with eigenvalue $\l$.  $\d^{(2)}$, $\d^{(3)}$, $\cdots$, can be solved
from \rf{alg} uniquely, except for $su(6p)$ and $sp(6p)$, where
additional degeneracy exists. In \ct{ACFGZ} it is shown that
$\d_j^{(n)}$ vanishes if $ n > \k l_j $ for some integer $\k$, which
is less than or equal to the degeneracy of the eigenvalue $\l$, except
for the special cases of $su(6p)$ and $sp(6p)$ noted above.

Since the equation \rf{pde} for $\t_j^{(1)}$ is linear and
homogeneous, one can superpose the single--soliton solutions to get
general N--soliton solutions, at least up to the first order:
\br
   \t_j^{(1)} = \eps_1 \d_{1,j}^{(1)} \e^{\G_1} + \cdots +
    \eps_N \d_{N,j}^{(1)} \e^{\G_N},  \lab{firstorder}
\er
where $\eps_a$ for a = 1, $\cdots$, N are some arbitrary expansion
parameters, $\d_{a,j}^{(1)}$ is the first order coefficient of the
$a^{\rm th}$ soliton and is an eigenvector of the L matrix, as
discussed above.  For a general N--soliton solution, we use the
following expansion for $\t_j$:
\be
  \t_j = \sum_{\{m_a\}} \eps_1^{ m_1} \cdots \eps_N^{ m_N} \;
   \d_j(m_1, m_2, \cdots, m_N)\,
   \exp\(m_1\G_1+m_2\G_2  + \cdots + m_N\G_N\), \lab{expansion}
\ee
where
     $\d_j (0, 0, \cdots, 0) = 1$, and from \rf{firstorder}
\be
    \d_j(0,\cdots,0,m_a=1, 0,\cdots, 0) = \d_{a,j}^{(1)}.  \lab{init}
\ee
{}From now on, we will let $\epsilon_a$ = 1.

Let us  define
\br
   f(m_1,m_2,\cdots, m_N)& \equiv&
    {1\over m^2}\( (m_1\g_1 + \cdots + m_N\g_N )^2
    - (m_1 \g_1 v_1 + \cdots + m_N \g_N v_N )^2 \) \cr
    & = &\sum_a m_a^2 \l_{s_a}
      + 2 \sum_{a \not= b} m_a m_b \sqrt{\l_{s_a} \l_{s_b} }
    \cosh \th_{ab}, \lab{deff}
\er
where $\th_{ab}$ is the rapidity difference of $a^{\rm th}$ and $b^{\rm th}$
soliton,  and $\l^{(a)}$ is the eigenvalue for the $a^{\rm th}$ soliton,
we have assumed that all the $\g$'s are positive just for definiteness.
Using the the following property of Hirota's derivative operators:
\be
  (D_t^2-D_x^2)\, \e^{\a_1 x + \b_1 t} \cdot \e^{\a_2 x + \b_2 t}
  =\Bigl((\b_1 -\b_2)^2 -(\a_1 -\a_2)^2\Bigr) \e^{(\a_1+\a_2)x+(\b_1+\b_2)t},
\ee
we can write:
\br
   \lefteqn{ {1\over 2 m^2}(D_t^2-D_x^2)\,\tau_j \cdot \tau_j + l_j \t_j^2 =}
    \lab{A} \\
    && - \sum_{k_a} \biggl( (f(k_1, \cdots, k_N)-2l_j) \d_j(k_1,\cdots,k_N)\,
    + \, A_j(k_1,\cdots,k_N) \biggr) \e^{ k_1\G_1+\cdots+k_N\G_N}, \nonumber
\er
where
\br
  A_j(k_1,\cdots,k_N) =
    \sum_{0\prec(m_1,\cdots,m_N)\prec(k_1,\cdots,k_N)}  \nonu
   &&   \( \h   f(k_1-2m_1,\cdots,k_N-2m_N) - l_j\)\cdot \cr
   &&   \d_j(m_1,\cdots,m_N)  \d_j(k_1-m_1,\cdots,k_N-m_N).  \nonu
\er
The partial ordering $(m_1,\cdots,m_N)\prec(k_1,\cdots,k_N)$ is defined
as
   $m_a\leq k_a$, for all $a$'s, and $ \sum m_a < \sum k_a $.
Also we can write:
\be
   \prod_{k\not=j} \t_k^{-K_{jk}} = \sum_{k_1,\cdots,k_N}
    \biggl( -\sum_{k\not=j} K_{jk}\d_k(k_1,\cdots,k_N)\,
    + \, B_j(k_1,\cdots,k_N) \biggr)e^{ k_1\G_1+\cdots+k_N\G_N},  \lab{B}
\ee
where $B_j(k_1,\cdots,k_N)$ is a function of the
$\d_j(m_1,\cdots,m_N)$'s for $(m_1,\cdots,m_N)\prec(k_1,\cdots,k_N)$.
The explicit form of $B_j(k_1,\cdots,k_N)$ will depend on the Cartan
matrix K.  Using \rf{A} and \rf{B}, the nonlinear partial differential
equation \rf{eom3} can now be simplified to a set of linear algebraic
equations:
\be
   \sum_k \Bigl(L_{jk}- f(k_1,\cdots,k_N) \d_{jk} \Bigr) \d_k(k_1,\cdots,k_N)
    = V_j (k_1,\cdots,k_N),
  \lab{*}
\ee
where $V_j (k_1,\cdots,k_N) = A_j(k_1,\cdots,k_N) + l_j B_j(k_1,\cdots,k_N)$.
Since the  r.h.s.\ of \rf{*} only depends on the preceding $\d_j$'s,
as discussed after \rf{B}, the above equation can  be solved iteratively.

First we decompose the right hand side of \rf{*} into linear
combinations of the eigenvectors,
\be
   V_j (k_1,\cdots,k_N) = \sum_s c_s(k_1,\cdots,k_N) \L_{s,j}, \lab{decomp}
\ee
where $\L_s$, for s = 0, 1, $\cdots$, r, is the complete set of
eigenvectors of the matrix L.  Since L is not necessarily symmetric,
the eigenvectors are not necessarily orthogonal to each other with
respect to the usual inner product. But for simply--laced loop
algebras, eigenvectors of the matrix L corresponding to distinct
eigenvalues are orthogonal to each other with respect to the diagonal
metric $(1/l_0, 1/l_1, \cdots, 1/l_N)$.  This observation will make
the decomposition \rf{decomp} easier.  Now from equations \rf{*} and
\rf{decomp} one can see that
\be
  \d_j(k_1,\cdots, k_N) =
   \sum_s {c_s(k_1,\cdots,k_N) \L_{s,j} \over \l_s-f(k_1,\cdots,k_N)}.
  \lab{dd}
\ee

Because of the iterative nature of \rf{*}, and from the
initial choice  \rf{init}, one can  show that
\br
  && \d_j(0,\cdots,0,m_a,  0,\cdots, 0) = \d_{a,j}(m_a), \cr
  && \d_j(0,\cdots,0,m_a,  0,\cdots,m_b,\cdots, 0)
      = \d_{ab,j}(m_a,m_b), \lab{*2}\\
  && \hskip 50pt  \vdots  \nonu
\er
where the $\d_{ab,j}(m_a,m_b)$ is a coefficient from the pair
scattering solution of the $a^{\rm th}$ and $b^{\rm th}$ soliton, etc.
This means that the coefficients of $(N-1)$--soliton solutions can be
used in the N--soliton solutions, this property makes solving general
N--soliton solutions tractable.

We conclude this section with a number of observations and remarks,
the first of which concerns integrability.  Due to it, the
perturbative expansion terminates at finite order. By counting the
powers in $\epsilon$, it is shown in \cite{ACFGZ} that, for a
single--soliton solution, the highest nonvanishing order in $\t_j$ is
$\k l_j$ for some constant integer $\k$.  Assuming that the expansion
\rf{expansion} gives exact solutions, then by repeating the power
counting for each $\eps_a$, one can show that:
\be
   \d_j(m_1,\cdots, m_N) = 0, \qquad \mbox{if} \quad
    (m_1,\cdots,m_N) \not\preceq (\k_1' l_j, \cdots, \k_N' l_j). \lab{*3}
\ee
For all the double--soliton solutions we obtained, we can explicitly
verify that $\k_a'=\k_a$ for $a = 1, 2$, where $\k_a$ is the $\k$
value for the $a^{\rm th}$ soliton.  For a general N--soliton solution
it is also true that $\k_a'=\k_a$ for $a = 1, 2, \cdots, N$, since
otherwise the resulting expression would not reduce to N
single--soliton solutions in the asymptotic limit.  Let $l_{max}={\rm
max}(l_1, \cdots, l_N)$, from the above discussion and \rf{*} we get:
\be
   V_j(m_1,\cdots,m_N)=0, \qquad \mbox{if} \quad
    (m_1,\cdots,m_N) \not\preceq l_{max} \; (\k_1, \cdots, \k_N). \lab{VV}
\ee
We consider a solution found if all nonvanishing $\d_j$'s are found.

If the Dynkin diagram of the loop algebra $\hat g_1$ has non--trivial
automorphisms, then $\hat g_1$ can be twisted to form a new loop
algebra $\hat g_2$.  The equations of motion of the Toda theory
associated to $\hat g_1$ will also have related symmetries, and thus
can be reduced to the equations of motion of the Toda theory
associated to $\hat g_2$ by identifying certain fields.  Also,
solutions to the $\hat g_1$ theory may be reduced (``folded'') to
solutions to the $\hat g_2$ theory \ct{mm,otfold}.  Furthermore, every
solution to the $\hat g_2$ theory can be unfolded to a solution to the
$\hat g_1$ theory, so we only need to deal with simply--laced loop
algebras.

Finally, we want to point out how plane--wave solutions come out of
the formalism.  If we replace the single--soliton ansatz \rf{ansatz}
by the plane--wave ansatz
\be
   \t_j^{(n)} = \d_j^{(n)}\,\e^{\i n\Phi},
\ee
where $\Phi = (k x - \omega t) + \xi$, and $\o^2-k^2 = m^2\l$, most of
the discussion on the solitons is still valid for the plane waves.
Even though we will only write down explicitly the soliton---soliton
scattering solutions, soliton---plane--wave and wave--wave scattering
solutions can be easily derived by changing some parameters from real
to purely imaginary.

%%%%%%%%%%%%%%%%%%%%%%%%%%%%%%%%%%%%%%%%%%%%%%%%%%%%%%%%%%%%%%%%%%%%%%%%%%
\section{$\hat A_r$ theory}
\setcounter{equation}{0}
The Dynkin diagram for $A_r^{(1)}$ is shown in Figure 3.1.
\bigbreak
\centerline{
\begin{picture}(250,110)(0,-10)
\put ( 18,50){\line( 1, 0){24}}
\put ( 58,50){\line( 1, 0){24}}
\put ( 98,50){\line( 1, 0){14}}
\put (168,50){\line( 1, 0){24}}
\put (208,50){\line( 1, 0){24}}
\put (152,50){\line(-1, 0){14}}
\put (132,95){\line(5,-2){103}}
\put (118,95){\line(-5,-2){103}}
\put ( 10,50){\circle{10}}
\put ( 50,50){\circle{10}}
\put ( 90,50){\circle{10}}
\put (160,50){\circle{10}}
\put (200,50){\circle{10}}
\put (240,50){\circle{10}}
\put (125,98){\circle{10}}
\put (10,35){\makebox(0,0){$\alpha_1$}}
\put (50,35){\makebox(0,0){$\alpha_2$}}
\put (90,35){\makebox(0,0){$\alpha_3$}}
\put (160,35){\makebox(0,0){$\alpha_{r-2}$}}
\put (200,35){\makebox(0,0){$\alpha_{r-1}$}}
\put (240,35){\makebox(0,0){$\alpha_r$}}
\put (125,114){\makebox(0,0){$\alpha_0$}}
\put (125,50){\makebox(0,0){.....}}
\put (125,10) {\makebox(0,0){Figure 3.1: Affine Dynkin diagram for
 $\hat A_r$.}}
\end{picture}
}

Since $l_j$ = 1, for all j's,  so $L_{jk}=K_{jk}$,
and Hirota's equation \rf{eom3} reads
\be
  {1\over 2m^2}(D_t^2 - D_x^2)\,\tau_j \cdot \tau_j   =  \(
  \tau_{j+1} \tau_{j-1}- \tau_j^{2} \), \qquad \mbox{for $j=0,1,2,\cdots,r.$}
  \lab{eomA}
\ee
Due to the periodicity of the extended Dynkin diagram, it is implied
that $\tau_{j+r+1} = \tau_{j}$.  The explicit expression for
$B_j(k_1,\cdots,k_N)$ is given below:
\br
  B_j(k_1,\cdots,k_N) =
    \sum_{0\prec(m_1,\cdots,m_N)\prec(k_1,\cdots,k_N)}
  \d_{j+1}(m_1,\cdots,m_N)  \d_{j-1}(k_1-m_1,\cdots,k_N-m_N).  \lab{B1}
\er
The matrix L has eigenvalues and eigenvectors
\be
   \l_s =  4 \sin^2 \vth_s, \quad
   \L_s = \(1,\;\e^{2\i \vth_s},\; \e^{4\i \vth_s}, \cdots,
    \; \e^{2\i r \vth_s} \),  \quad \mbox{for s=1, 2, $\cdots$, r},
   \lab{eigenA}
\ee
where $\vth_s = s \pi /h $, and h = $r+1$ is the Coxter number.

For any given integer $1 \leq s \leq r$, there is a class of
single--soliton solutions ($\k=1$) given in \ct{hollo1}, with the only
non--vanishing coefficients $\d^{(1)} = \L_{s}$. The topological
charge of the soliton depends on the value of the parameter $\xi$.
Because of degeneracy in eigenvalues, $\l_s = \l_{h-s}$, there also
exist $\k=2$ solitons \ct{ACFGZ}:
\be
   \d^{(1)}_{j} = y^+ \exp\({2\pi \i sj /h}\) + y^- \exp\(-{2\pi \i sj/h}\),
   \qquad  \d_{s,j}^{(2)} = y^+ y^- \cos^2 \vth_s. \lab{2single}
\ee
Here $y^+$ and $y^-$ are independent variables, if we take either of
$y^+$ and $y^-$ be zero in the above equations, we will get back to
$\k=1$ solutions.  In the following we will construct multi--soliton
solutions consisting of solitons with $\k=2$, since if there are some
$\k=1$ solitons involved, we only need to take some of the $y$'s to be
zero.

First we solve for a two--soliton solution.  Let $\l^{(1)} = \l_p$,
$\l^{(2)} = \l_q$, p, q are integers with $1\le p,\ q \le r$, and
$\k_1 = \k_2 =2$, from \rf{firstorder}, \rf{*2} and \rf{2single},
\br
   \d_j(1,0) &=&y^+_1 \exp\({2\pi \i pj /h}\)+y^-_1 \exp\(-{2\pi \i pj/h}\),\cr
   \d_j(0,1) &=&y^+_2 \exp\({2\pi \i qj /h}\)+y^-_2 \exp\(-{2\pi \i qj/h}\),\\
   \d_j(2,0)&=& y^+_1 y^-_1 \cos^2 \vth_p, \qquad
   \d_j(0,2) = y^+_2 y^-_2 \cos^2 \vth_q.   \nonumber
\er
By straightforward calculations, we get:
\br
   \d_j(1,1) &=& y^+_1 y^+_2 \, \e^{\g_{pq}^{++}}\, \exp\({2\pi \i (p+q)j/h}\)
          + y^-_1 y^-_2 \, \e^{\g_{pq}^{--}}\, \exp\(-{2\pi \i (p+q)j/h}\)\cr
          &  +& y^+_1 y^-_2 \, \e^{\g_{pq}^{+-}}\, \exp\( {2\pi \i (p-q)j/h}\)
          + y^-_1 y^+_2 \, \e^{\g_{pq}^{-+}}\, \exp\(-{2\pi \i (p-q)j/h}\),\cr
   \d_j(2,1) &=& \d_j(2,0) \d_j(0,1) \exp({\g_{pq}^{++} + \g_{pq}^{+-} }),\\
   \d_j(1,2) &=& \d_j(0,2) \d_j(1,0) \exp({\g_{pq}^{++} + \g_{pq}^{+-} }),\cr
   \d_j(2,2) &=& \d_j(2,0) \d_j(0,2) \exp({2\g_{pq}^{++} + 2\g_{pq}^{+-} }),
        \nonumber
\er
and all higher order terms vanish.  Here
\br
  \exp({\g_{s_a,s_b}^{++}}) = \exp({\g_{s_a,s_b}^{--}})
   = \exp({\g_{s_a,s_b}^{+}})
   ={\cosh \th_{ab} - \cos \vth_{s_a-s_b} \over \cosh \th_{ab} -
   \cos\vth_{s_a+s_b} }, \lab{g+} \\
  \exp({\g_{s_a,s_b}^{+-}}) = \exp({\g_{s_a,s_b}^{-+}})
   = \exp({\g_{s_a,s_b}^{-}})
   ={\cosh \th_{ab} + \cos \vth_{s_a+s_b} \over \cosh \th_{ab} +
   \cos \vth_{s_a-s_b} }. \lab{g-}
\er

Constructing an N--soliton solution means finding all nonvanishing
$\d_j(m_1,\cdots, m_N)$.  From \rf{*2}, if any of the $m_a$ vanishes,
then the corresponding $\d$ can be taken from some $N-1$ soliton
solutions.  Since we have assumed $\k=2$, from \rf{*3}, if any of the
$m_a$ is greater than or equal to 3, then the corresponding $\d$ shall
vanish.  So we only need to consider $1 \leq m_a \leq 2$.  Furthermore
from the above two--soliton solutions we see that $\d_j(2,1)$ and
$\d_j(1,2)$ are related by renaming the solitons, or shuffling the
indices $p \leftrightarrow q$. This is also true for N--soliton
solutions, so we only have to find
$\d_j(\overbrace{2,\cdots,2}^{i},\overbrace{1,\cdots,1}^{N-i} )$,
for i = 0, 1, $\cdots$, N.

Let $\l^{(a)} = \l_{s_a}$, with $1\le s_a\le r$, We list the
nontrivial coeffients for N--soliton solutions below:
\def\mv{&&\kern -50pt}
\br
  \mv \d_j(\overbrace{1,\cdots,1}^{N}) = \sum_{\{\s\}} y_1^{\s_1} \cdots
y_N^{\s_N}
      \exp{ \(i 2 \pi(\s_1 s_1 + \cdots + \s_N s_N)j /h \)} \exp\Bigl(
      \sum_{1\leq b<c \leq N} \g_{s_b, s_c}^{\s_b \s_c}\Bigr),    \lab{a111} \\
  \mv \d_j(\overbrace{2,\cdots,2}^{i},\overbrace{1,\cdots,1}^{N-i})
      = \d_j(\overbrace{2,\cdots,2}^{i},\overbrace{0,\cdots,0}^{N-i})\;
      \d_j(\overbrace{0,\cdots,0}^{i},\overbrace{1,\cdots,1}^{N-i})
      \prod_{a \leq i \atop b > i} \exp{(\g_{s_a, s_b}^+ + \g_{s_a, s_b}^-) },
      \lab{a221}  \\
  \mv \d_j(\overbrace{2,\cdots,2}^N) = \d_{1,j}^{(2)} \cdots \d_{N,j}^{(2)}
      \prod_{a<b} \exp{(2\g_{s_a, s_b}^+ + 2\g_{s_a, s_b}^-) }.     \lab{a222}
\er

We will outline our construction and notation in three steps, assuming
all the $N-1$ soliton solutions are known.

Step 1:  To get \rf{a111}.

One can easily see that $\d_j(\overbrace{1,\cdots,1}^{N})$ should be
N--linear in the $y$'s. Note that $y_1^+ y_1^-$ is  considered to be
quadratic while $y_1^+ y_2^-$ is bilinear.    So
   $$\d_j(\overbrace{1,\cdots,1}^{N}) = \sum_{\{\s\}} C(\s_1,\cdots,\s_N)
     y_1^{\s_1} \cdots y_N^{\s_N},  $$
where $\s_a=\pm 1$ and $C(\s_1,\cdots,\s_N)$ are independent of $x$, $t$
and all the $y$'s.  $y^\s$ should be understood as $y^+$ or $y^-$,
Summation is over all possible combinations of $\s$'s

If we take $y_1^-=\cdots=y_{N}^- = 0$, then only one term survives,
and all the $\k$'s become 1 instead of 2, so we should get those of
Hollowood,          \foot{In  \ct{hollo1} the N--soliton solutions
were written down without explicit proof.}
i.e.,
\be
   C(1,\cdots,1)=\exp\Bigl(\sum_{1\le b <c \le s_1} \g_{s_bs_c}^{+}\Bigr)
      \exp{ \(\i 2 \pi(s_1+\cdots+s_{N})j /h \)}. \lab{holsol1}
\ee
Proving  the above equation is equivalent to proving the following identity:
\be
  \sum_{\{ \s \} } \Bigl( f(\s_1, \s_2, \cdots, \s_N)
   - 4 \sin^2{(\s_1 \vth_1 + \cdots + \s_N \vth_N ) }\Bigr)
   \prod_{a < b } \Bigl(\cosh \th_{ab} - \cos {(\s_a \vth_a
  - \s_b \vth_b)}\Bigr)  =0,  \lab{identity}
\ee
where each $\s$ takes value 1 or $-1$, and summation is over all possible
combinations of $\s$.

Since directly proving the above identity is hard, we will construct
\rf{holsol1} without direct proof of \rf{identity}, by using \rf{VV} instead.
Since we have taken $y_1^-=\cdots=y_{N}^- = 0$, all the single solitons
have $\k = 1$.
By \rf{*3},  the only nonvanishing terms in
$V_j(1,\overbrace{2,\cdots,2}^{N-1})$ are
\br
   && \d_{j-1}(1,1,\cdots,1)\, \d_{j+1}(0,1,\cdots,1) +
    \d_{j+1}(1,1,\cdots,1)\, \d_{j-1}(0,1,\cdots,1) \cr
   && -2 \cos (2\pi s_1/h)\, \d_j (1,1,\cdots,1)\, \d_{j}(0,1,\cdots,1). \nonu
\er
Let $\d_j (1,1,\cdots,1) = \e^{\i 2\pi j(s_2+\cdots+s_N)/h} T_j$, from the
vanishing  condition of $V_j(1,\overbrace{2,\cdots,2}^{N-1})$ we get:
\be
  T_{j-1} + T_{j+1} - 2 \cos (2 \pi s_1/h) T_j = 0.
\ee
The above homogeneous second order difference equation has two linearly
independent solutions:
  $$  T_j = T \e^{\pm \i 2 \pi j s_1/h }.    $$
{}From a symmetry argument or the vanishing conditions $V_j(2,\cdots,2,1)=0$,
we can see that we should chose the ``+'' sign in the above solution.
The constant factor $T$ can  be determined from the vanishing condition of
$V_j(1,1, \overbrace{2,\cdots,2}^{N-2})$, which gives \rf{holsol1}.
We could find all the $C(\s_1, \cdots, \s_N)$'s the same way, thus proving
\rf{a111}.

Step 2:   To get \rf{a221}.

First let $y_{i+1}^-=\cdots=y_{N}^-=0$, so $\k_1=\cdots=\k_i=2, \quad
\k_{i+1}=\cdots=\k_{N}=1 $.  Adding all the nonvanishing terms in
$V_j(\overbrace{4,\cdots,4}^i,\overbrace{2,\cdots,2}^{N-i-1},1)$, from
Eq.\ \rf{VV}, the sum should vanish. So we obtain a homogeneous second
order difference equation of
$\d_j(\overbrace{2,\cdots,2}^i,1,\cdots,1)$.  We can write down the
vanishing condition of
$V_j(\overbrace{0,\cdots,0}^i,\overbrace{2,\cdots,2}^{N-i-1},1)$ also,
which gives a homogeneous second order difference equation of
$\d_j(\overbrace{0,\cdots,0}^i,1,\cdots,1)$.  Comparing the two
equations, and from the fact that they obey the same boundary
conditions and have similar symmetries, we see that
$\d_j(\overbrace{2,\cdots,2}^i,1,\cdots,1)$ should be proportional to
$\d_j(\overbrace{0,\cdots,0}^i,1,\cdots,1)$. The proportionality
constant can be determined from the vanishing condition:
$$V_j(\overbrace{4,\cdots,4}^i,\overbrace{2,\cdots,2}^{N-i-2},1,1)=0.$$

Second, there should be $2^{N-i}$  terms, and each can be derived
in a similar way. Putting them all together, we will get \rf{a221}.

Step 3:  To get \rf{a222}.

{}From the vanishing condition of
$V(4,\cdots,4,3)$ and using factorization of  \rf{a221}, we obtain:
\be
   (\l_N -2) \d_j(2,\cdots,2) + \e^{\pm \i 2\pi s_N/h} \d_{j-1}(2,\cdots,2)
    + \e^{\mp \i 2\pi s_N/h} \d_{j+1}(2,\cdots,2) = 0.
\ee
Adding and subtracting the two equations, we see that $\d_j(2,\cdots,2)$
is constant, which constant can be calculated from the vanishing
conditions of  $V(4,\cdots,4,3,3)$.

\vskip 4mm
This completes the construction.  These $\kappa = 2$ solitons are not
really ``single solitons'' in the strict sense, since they change
shape after scattering with other ($\k = 1$) solitons. In general a
$\kappa = 2$ soliton has two peaks, and each peak behaves as if it is
a free soliton when scattering off other solitons.  These $\k=2$
solitons should be interpreted as compounds of two $\k=1$ solitons
interacting in a peculiar way so that there seems to be no interaction
among the peaks.  Indeed, one can obtain such $\k=2$ soliton solutions
by taking the static limit of some two--soliton scattering solutions
of $\k=1$ solitons \ct{mcghee}.

Let us look into the scattering of two $\k=1$ solitons in a little
more detail.  Since the only effect after scattering is a shift in the
center of mass for each soliton, and the shift does not depend
directly on the topological charge of each soliton but rather on the
topological classes they belong to, one may be tempted to believe that
the interations among the solitons depend only on the classes.
Furthermore, since $\g_{ab}^+$ in \rf{g+} is always negative, it
appears as if the interations are always attractive.  However, on
closer inspection, one notices that the two solitons could either
speed up or slow down as they approach each other.  If they speed up,
eventually the two peaks will merge together and then separate again.
If, on the other hand, they slow down, as they approach to within a
certain distance, they begin to exchange shape, topological charge and
other attributes, and then move apart again as if the two solitons
have switched position. Because the two solitons performed this
exchange over some finite distance, even though the interaction is
repulsive, the total shift looks like an attractive case.  In both
cases, there is no reflection even if the masses are equal.  Details
of which pair of solitons attract or repel each other still need to be
worked out.  Since we can also have two solitons stand still as if
there is no interaction at all, the interaction cannot be described by
a simple equivalent potential. This may be because the solitons are
extended objects.

If $r = 2 N -1$, for some integer N = 2, 3, $\cdots$, and
\be
   \t_j = \t_{r-j},  \quad \mbox{for j = 1, 2, $\cdots$, N,} \lab{symmetry}
\ee
then the set {$\t_0,\ \t_1,\ \cdots,\ \t_N$} will form a tau-function
solution set for sp(2N) affine Toda models.  It is easy to check that
if we set $y^+ = y^-$, then the $\k = 2$ soliton given in \rf{2single}
will satisfy the condition \rf{symmetry}. These $\k=2$ solitons of
sp(2N) do look and behave like single solitons: the two overlapping
peaks do not drift apart since they are protected by the symmetry
\rf{symmetry}.  From the solutions  \rf{a111} to \rf{a222} one can also
easily obtain multi--soliton solutions for sp(2N) affine TFTs.

%%%%%%%%%%%%%%%%%%%%%%%%%%%%%%%%%%%%%%%%%%%%%%%%%%%%%%%%%%%%%%%%%%%%%%%%%%%%
\section{$\hat D_4$}
Dynkin diagram for $\hat D_4$ is shown below:

\bigbreak
\centerline{
\begin{picture}(100,120)(0,5)
\put ( 45, 85){\line(-1,-1){20}}
\put ( 55, 85){\line( 1,-1){20}}
\put ( 55, 95){\line( 1, 1){20}}
\put ( 45, 95){\line(-1, 1){20}}
\put ( 20, 60){\circle{10}}
\put ( 50, 90){\circle{10}}
\put ( 80, 60){\circle{10}}
\put ( 20,120){\circle{10}}
\put ( 80,120){\circle{10}}
\put ( 20,105){\makebox(0,0){$\alpha_0$}}
\put ( 20, 45){\makebox(0,0){$\alpha_1$}}
\put ( 50, 75){\makebox(0,0){$\alpha_2$}}
\put ( 80,105){\makebox(0,0){$\alpha_3$}}
\put ( 80, 45){\makebox(0,0){$\alpha_4$}}
\put ( 50, 20){\makebox(0,0){Figure 5.1: Affine Dynkin diagram for
$\hat D_4$.}}
\end{picture}
}

$l_0=l_1=l_3=l_4=1, \qquad l_2=2$.
\br
   \mv B_0(k_1,\cdots, k_N)=B_1(k_1,\cdots, k_N)
     = B_3(k_1,\cdots, k_N) = B_4(k_1,\cdots, k_N) =0, \cr
   \mv B_2(k_1,\cdots, k_N) = \sum
     \d_0(m_1,\ldots,m_N)  \d_1(n_1,\ldots,n_N) \d_3(p_1,\ldots,p_N)
     \d_4(q_1,\ldots,q_N), \nonu
\er
where the summation is over ${p_a+q_a+m_a+n_a=k_a}$.
The eigenvalues and eigenvectors of the matrix $L_{jk}$ are:
\br
  & \l_1 = 2;  \qquad  & \L_1 = (1,\; -1,\; 0,\; 1,\; -1),  \cr
  & \l_2 = 2;  \qquad  & \L_2 = (1,\; 1,\; 0,\; -1,\; -1),  \cr
  & \l_3 = 2;  \qquad  & \L_3 = (1,\; -1,\; 0,\; -1,\; 1),  \cr
  & \l_4 = 6;  \qquad  & \L_4 = (1,\; 1,\; -4,\; 1,\;  1),  \nonu
\er
We also define  $\L_6 \equiv (0,\; 0,\; 1,\; 0,\; 0) $ for later convenience.

For $\l = 6$, the eigenvalue is not degenerate, there is only one corresponding
single soliton solution with $\k = 1$:
  $$ \d^{(1)} = \L_4; \qquad \d^{(2)} = \L_6. $$
However, the eigenvalue $\l = 2$ has degeneracy 3, so one can construct a
$\k = 3$ soliton:
\br
  && \d^{(1)} = y_1 \L_1 + y_2 \L_2 + y_3 \L_3, \cr
  && \d^{(2)} = {1\over 3}( y_1 y_2 \L_3 + y_2 y_3 \L_1 + y_1 y_3 \L_2)
       + (y^2_1 + y^2_2 + y^2_3)\L_6, \cr
  && \d^{(3)} = P_3 (\L_4 - 12 \L_6), \qquad \d^{(4)} = P_4 \L_6, \lab{k3} \\
  && \d^{(5)} = 0, \qquad \d^{(6)} = P_3^2 \L_6, \nonu
\er
where $P_3 = {y_1 y_2 y_3 \slash 27},\quad
   P_4 = {y_1^2 y_2^2 + y_1^2 y_3^2 + y_2^2 y_3^2 \slash 9}$.
One can obtain three $\k = 2$ or $\k = 1$ solitons if one or two of the y's
is taken to be zero.

We will first study the two--soliton scattering solutions. Denote:
\be
   C_1 = {\cosh \theta - 1 \over \cosh \theta +1 },  \quad
   C_2 = {  2\cosh \theta -1 \over  2\cosh \theta + 1 }, \quad
   C_3 = {  2\cosh \th -\sqrt{3} \over 2\cosh \th + \sqrt{3} }, \quad
   C_4 = {  2\cosh \theta +1 \over  2\cosh \theta - 1 }.
\ee
Since it is straight forward to calculate the double--soliton solutions,
we will write down the nontrivial coefficients without derivation.

For  $\l^{(1)}=\l^{(2)}=6$:
\br
   && \d (1,0) = \d (0,1) = \L_4, \quad\qquad \d (2,0) = \d (0,2) = \L_6, \cr
   && \d (1,1) = {2(5 \cosh \th - 4) \over 3(\cosh \th +1)} \L_0 +
      {(11 - 14\cosh \th) \over 3(2 \cosh \th +1)} \L_4, \lab{c66}\\
   && \d (1,2) = \d (2,1) = -4 C_1 C_2 \L_6, \qquad
      \d (2,2) = C_1^2 C_2^2 \L_6. \nonu
\er

For $\l^{(1)}=2$ with $\k = 3$, and $\l^{(2)}=6$ with $\k=1$:
$\d(m,0)$ will be taken as $\d^{(m)}$ in \rf{k3}, $\quad \d(0,1) = \L_4,
\quad \d(0,2) = \L_6$, and
\br
   \d (1,1) &=& C_3 ( y_1 \L_1 + y_2 \L_3 + y_3 \L_3 ), \cr
   \d (2,1) &=& {1\over 3}C_3^2 ( y_1 y_2 \L_3 + y_2 y_3 \L_1 + y_1 y_3 \L_2 )
     - 4 C_3 ( y_1^2 + y_2^2 + y_3^2 )  \L_6,   \cr
   \d (2,2) &=& C_3^2 ( y_1^2 + y_2^2 + y_3^2 ) \L_6, \cr
   \d (3,1) &=&  P_3 C_3^3 \L_4 + 4 P_3 (9 C_3^2 +9 C_3 -1) \L_6, \lab{c26}\\
   \d (3,2) &=& 16 P_3  C_3^3 \L_6,  \cr
   \d (4,1) &=& -4 P_4 C_3^2 \L_6, \qquad \d (4,2) = P_4 C_3^4  \L_6,  \cr
   \d (6,1) &=& - 4 P_3^2 C_3^3 \L_6, \qquad \d (6,2) = P_3^2 C_3^6 \L_6. \nonu
\er

For $\l^{(1)}= \l^{(2)}=2$ with $\k_1 =3,\quad \k_2 = 1$:
Let  $\d(m,0)$ will be taken as $\d^{(m)}$ in \rf{k3}, $\quad \d(0,1) = \L_1,
\quad \d(0,2) = \L_6$, and
\br
  \d (1,1) &=& y_1 \( C_1 C_4 \L_0 - {2(2 \cosh \th - 3) \over
    2 \cosh\th - 1} \L_6 \)  + C_2 (y_2 \L_3 + y_3 \L_2 ),  \cr
  \d (2,1) &=& y_2 y_3\( {1\over 3} C_2^2 \L_4 + {4(2 \cosh \th -3) \over
     3 (2 \cosh \th + 1)} \L_6 \) +
    {1 \over  3} C_1 y_1 ( y_2 \L_2 + y_3 \L_3 ), \cr
  \d (2,2) &=& ( y_1^2 C_1^2 C_4^2 + y_2^2 C_2^2 + y_3^2 C_2^2) \L_6, \cr
  \d (3,1) &=& {1\over 27} y_1 y_2 y_3 C_1 C_2 \L_1 + {4 y_1 (y_2^2 + y_3^3)
     \over 3 (1 + \cosh \th) (1 + 2 \cosh \th) } \L_6,    \lab{c22}\\
  \d (3,2) &=& -{16 \over 27} C_1 C_2 y_1 y_2 y_3 \L_6,    \cr
  \d (4,1) &=& {16  \over 27} { C_1 \over (1 - 2 \cosh \th)(1 + 2 \cosh \th)}
    y_1 y_2 y_3 \L_6,   \cr
  \d (4,2) &=& {1 \over 9} \Bigl( C_1^2 y_1^2 (y_2^2 + y_3^2)
   + C_2^4 y_2^2 y_3^2 \Bigr) \L_6,   \cr
  \d (5,1) &=& {4 \over 81} {C_2 y_1 y_2^2 y_3^2 \over (1 + \cosh \th)
    (1 + 2 \cosh \th)}  \L_6,   \cr
  \d (6,2) &=& {1 \over 729} C_1^2 C_2^2 y_1^2 y_2^2 y_3^2 \L_6. \nonu
\er

{}From the solutions \rf{c66} and \rf{c26}, with a little analysis, we can see
that the  solitons retain their shape after scattering.  However in
the solution \rf{c22}, the  $\k = 3$ soliton Changes shape after scattering,
as signified by the fact that the constant $y_1$ is treated differently from
$y_2$ and $y_3$.
The $\k = 3$ soliton can be thought of as a composite of three
$\k =1$ solitons. We will substantiate this claim by writing down the
three--soliton scattering solution.
Let $\d(1,0,0) = y_1 \L_1$, $\d(0,1,0) = y_2 \L_2$,
    $\d(0,0,1) = y_3 \L_3$. Then $\d(2,0,0) = y_1^2 \L_6$,
    $\d(0,2,0) = y_2^2 \L_6$,  $\d(0,0,2) = y_3^2 \L_6$,
and
\br
   && \d(1,1,0) = y_1 y_2 D_{12} \L_3, \quad \d(1,0,1) = y_1 y_3 D_{13} \L_2,
      \quad \d(0,1,1) = y_2 y_3 D_{23} \L_1,\cr
   && \d(1,1,1) = y_1 y_2 y_3 D_{12} D_{23} D_{13} \L_0 -
      y_1 y_2 y_3 \( 2 D_{12} D_{23} D_{13} + 16 D_{123} \) \L_6, \lab{c222}\\
   && \d(2,2,0) = y_1^2 y_2^2 D_{12}^2 \L_6, \quad
      \d(2,0,2) = y_1^2 y_3^2 D_{13}^2 \L_6, \quad
      \d(0,2,2) = y_2^2 y_3^2 D_{23}^2 \L_6, \cr
   && \d(2,2,2) = y_1^2 y_2^2 y_3^2  D_{12}^2 D_{23}^2 D_{13}^2 \L_6,\nonu
\er
where $ D_{ab}={\displaystyle{2 \cosh \th_{ab} -1 \over 2 \cosh
\th_{ab} +1}}$ and $D_{123} = {\displaystyle{ 1 \over (2 \cosh
\th_{12} +1) (2 \cosh \th_{23} +1) (2 \cosh \th_{13} +1) }}$.  Now if
all the three--solitons have the same velocity, then $\th_{12} =
\th_{23} = \th_{13} =0$, and we can have $\G_1 = \G_2 = \G_3 = \G$.
Then the above three--soliton scattering solution will have the single
soliton form with $\d^{(m)} = \sum_{m_1 + m_2 + m_3 = m} \d(m_1, m_2,
m_3)$, which gives the solution set \rf{k3}.

Because of the non-trivial symmetries of the Dynkin diagram, the
solutions can be folded to yield solutions to $G_2$ affine TFT by
demanding $\t_1 = \t_3 = \t_4$. The $\k=3$ soliton with $y_1 = y_2 =
y_3$ and the $\l=6$ soliton survive the folding procedure. Protected
by the discrete symmetry, these will be true single solitons of $G_2$
affine TFT.

%%%%%%%%%%%%%%%%%%%%%%%%%%%%%%%%%%%%%%%%%%%%%%%%%%%%%%%%%%%%%%%%%%%%%%%%%%%%%
\section{$D_{2r}^{(1)}$ Theory for $r \geq 5$}
\setcounter{equation}{0}

The Dynkin diagram for $D_r^{(1)}$\ $(r\geq 5)$ is shown in Figure 4.1.
\bigbreak
\centerline{
\begin{picture}(250,120)(0,5)
     \put ( 45, 85){\line(-1,-1){20}}
     \put ( 58, 90){\line( 1, 0){24}}
     \put ( 98, 90){\line( 1, 0){14}}
     \put (168, 90){\line( 1, 0){24}}
     \put (205, 85){\line( 1,-1){20}}
     \put (152, 90){\line(-1, 0){14}}
     \put (205, 95){\line( 1, 1){20}}
     \put ( 45, 95){\line(-1, 1){20}}
     \put ( 20, 60){\circle{10}}
     \put ( 50, 90){\circle{10}}
     \put ( 90, 90){\circle{10}}
     \put (160, 90){\circle{10}}
     \put (200, 90){\circle{10}}
     \put (230, 60){\circle{10}}
     \put ( 20,120){\circle{10}}
     \put (230,120){\circle{10}}
     \put ( 20,105){\makebox(0,0){$\alpha_0$}}
     \put (238,105){\makebox(0,0){$\alpha_{r-1}$}}
     \put ( 20, 45){\makebox(0,0){$\alpha_1$}}
     \put ( 50, 75){\makebox(0,0){$\alpha_2$}}
     \put ( 90, 75){\makebox(0,0){$\alpha_3$}}
     \put (160, 75){\makebox(0,0){$\alpha_{r-3}$}}
     \put (200, 75){\makebox(0,0){$\alpha_{r-2}$}}
     \put (230, 45){\makebox(0,0){$\alpha_r$}}
     \put (125, 90){\makebox(0,0){.....}}
     \put (125, 20){\makebox(0,0){Figure 4.1: Affine Dynkin diagram for
      $D_r^{(1)}$.}}
\end{picture}
        }

$l_0=l_1=l_{r-1}=l_r = 1$, $\quad l_j=2$ \ \  for $2 \leq j \leq r-2$.
The explicit form of $B_j(k_1,\ldots,k_N)$ is   given below:
\br
   && B_0(k_1,\ldots,k_N) = B_1(k_1,\ldots,k_N) = B_{r-1}(k_1,\ldots,k_N)
    =   B_r(k_1,\ldots,k_N) = 0, \cr
   && B_2(k_1,\ldots,k_N) = \sum_{p_a + q_a + m_a = k_a}
    \d_0(p_1,\cdots,p_N) \d_1(q_1,\cdots, q_N) \d_3(m_1,\cdots, m_N), \\
   && B_{r-2}(k_1,\ldots,k_N) = \sum_{p_a + q_a + m_a = k_a}
    \d_r(p_1,\cdots,p_N) \d_{r-1}(q_1,\cdots, q_N) \d_{r-3}(m_1,\cdots, m_N),
    \nonu
\er
and the rest of the $B_j(k_1,\ldots,k_N)$'s are the same as in \rf{B1}.

The  matrix L has eigenvalues  and eigenvectors
\br
   && \l_s = 8\sin^2 \vth_s,  \qquad \hbox{for $ 1\leq s \leq r-2$};  \cr
   && \L_s = \Bigl(1,1, 2{\cos{3\vth_s}\over \cos{\vth_s}},
       2{\cos{5\vth_s}\over \cos{\vth_s}},
       \ldots, 2{\cos{(2r-5)\vth_s}\over \cos{\vth_s}}, (-1)^s,(-1)^s\Bigr),\\
   && \l_{r-1}=\l_r = 2;   \qquad
      \L_{r-1}= (1,-1,0,0,\ldots,0), \quad
      \L_r = (0,0,\ldots, 0, -1, 1), \nonu
\er
where $\vth_s={s\pi/2(r-1)}=s\pi/h$.

First we establish some useful identities for later calculations of
double--soliton solutions:
\br
   \sum_{j=2}^{r-2} \cos (2j-1)\vth_p = \cases {
     r-3, &if p=0 mod h; \cr
     -\Bigl( 1 + (-1)^p \Bigr) \cos \vth_p, & otherwise }
\er
\br
   \sum_{j=2}^{r-2} {\cos (2j-1)\vth_p \;\cos (2j-1)\vth_q \over
    \cos \vth_p \cos \vth_q}  = \cases { \displaystyle
    -1+{r-1 \over 2 \cos^2 \vth_p}, &if $p\pm q = 0$ mod h; \cr
    -\Bigl( 1 + (-1)^{p+q} \Bigr), &otherwise  }
\er
\br
   \lefteqn{
   \sum_{j=2}^{r-2} {\cos (2j-1)\vth_p \;\cos (2j-1)\vth_q\;\cos (2j-1)\vth_m
      \over\cos \vth_p \;\cos \vth_q\;\cos \vth_m } } \nonu\cr
   &=&\cases { \displaystyle
   -2+{r-1\over 4\cos \vth_p \;\cos \vth_q\;\cos \vth_m },
   &if $p\pm q\pm m = 0 $ mod h; \cr
   -\Bigl( 1 + (-1)^{p+q+m} \Bigr), &otherwise  }
\er
\br
   \lefteqn{
   \sum_{j=2}^{r-2} { \Bigl( \cos (2j-3)\vth_p \;\cos (2j+1)\vth_q
     + \cos (2j+1)\vth_p \;\cos (2j-3)\vth_q \Bigr) \cos (2j-1)\vth_m
     \over \cos \vth_p \;\cos \vth_q\;\cos \vth_m } }  \nonumber\\
   &=& \cases { \displaystyle
   -2 \( {\cos 3\vth_p \over \cos \vth_p }+{\cos 3\vth_p \over \cos \vth_p }\)
   +{(r-1) \cos 2 \vth_{p+q} \over \cos \vth_p \;\cos \vth_q\;\cos \vth_m },
     &if $m\pm (p-q) = 0 $ mod h; \nonu\cr
    \displaystyle
   -2 \( {\cos 3\vth_p \over \cos \vth_p }+{\cos 3\vth_p \over \cos \vth_p }\)
   +{(r-1) \cos 2 \vth_{p-q} \over \cos \vth_p \;\cos \vth_q\;\cos \vth_m },
     &if $m\pm (p+q) = 0 $ mod h;\nonu \cr
     \displaystyle
   -\Bigl( 1 + (-1)^{p+q+m} \Bigr) \( {\cos 3\vth_p \over \cos \vth_p }
   + {\cos 3\vth_p \over \cos \vth_p } \), &otherwise.    }
\er

Single--soliton solutions have been given in \cite{ACFGZ}.\ \  For
$1 \leq s \leq r-2$, the eigenvalues are non--degenerate, $\k=1$,
\br
   \d^{(1)} = \L_s, \qquad
   \d^{(2)} = u \equiv (0,\; 0,\;1,\;  \ldots,\; 1,\; 0,\; 0). \lab{b1}
\er
For the eigenvalue $\l=2$, which is doubly degenerate, we can have a $\k=2$
soliton:
\br
   && \d^{(1)} = y_1 \L_{r-1} + y_2 \L_r, \cr
   && \d^{(2)} = \Bigl(\; c, c, d_2, d_3, \cdots, d_{r-2},
       (-1)^{r-1}c, (-1)^{r-1}c \;\Bigr), \lab{b2}\\
   && \d^{(4)} = c^2 u = c^2 (0, 0, 1,1,\cdots,1,0,0 ), \nonu
\er
where $c = (y_1^2 + (-1)^{r-1} y_2^2)/2h$, and
$ d_j = (-1)^j  \Bigl( (h - 2j +1)y_1^2 + (-1)^r (2j - 1) y_2^2 \Bigr)/h $.
If we let  $y_2 = \pm (-1)^{r/2} y_1$, then the above $\k=2$ soliton will be
reduced to a  $\k=1$ soliton.

We will next write down some double--soliton solutions without
derivation.  Let $\l^{(1)} = \l_p$, and $\l^{(2)} = \l_q$, for some
integers $1 \leq p\, , q \leq r-2$, then $\d_j(1,0)= \L_{p,j}$,
$\d_j(0,1) = \L_{q,j} $, and
\br
   && \d_j(0,2) = \d_j(2,0) = u_j, \cr
   && \d_0(1,1) = \d_1(1,1) =  -1 + {\cos (\vth_p + \vth_q)\over
     \cos \vth_p\, \cos \vth_q} \e^{\g_{pq}^+}
     +{\cos (\vth_p - \vth_q)\over
     \cos \vth_p\,\cos \vth_q} \e^{\g_{pq}^-}, \cr
  && \d_r (1,1) = \d_{r-1}(1,1) =  {{( -1 ) }^{p + q}}\, \d_0(1,1), \cr
  && \d_j(1,1)  =  {2\cos ( 2\,j -1)\, \vth_{p+q}
     \over \cos \vth_p\,\cos \vth_q } \e^{\g_{pq}^+}
     +   { 2\,\cos (2\,j-1 ) \, \vth_{p - q }  \over
     \cos \vth_p\,\cos \vth_q }  \e^{\g_{pq}^-},  \quad
     \mbox{j=2, \ldots, r-2}, \\
  && \d_j(2,1) = u_j\, \d_j(1,0)\, \d_0(1,1), \cr
  && \d_j(1,2) = u_j\, \d_j(0,1)\, \d_0(1,1),\cr
  && \d_j(2,2) = u_j\, \Bigl( \d_0(1,1) \Bigr)^2.  \nonu
\er

Let  $\l^{(2)} = \l_p$ for some $0<p \leq r-2$, $\l^{(2)} = 2$,  and
$\d_j(1,0)= \L_{p,j}, \quad \d_j(0,1) =  \L_{r-1} \pm (-1)^{r/2} \L_r$,
then:
\br
  && \d (2,0)=u, \qquad \d (0,2) = \(0,0,1,-1,1,-1,\cdots,(-1)^{r-2},0,0\), \cr
  && \d (1,1) =  (\L_{r-1} \pm (-1)^{{r\over 2}+p} \L_r)\; \e^{\tilde \g}, \cr
  && \d_j(1,2)=(-1)^j \L_{p,j}\; \e^{\tilde \g}, \quad \mbox{for }
     2\leq j \leq r-2, \\
  && \d_j(2,2)= \d_j(0,2)\; \e^{2 \tilde \g}, \nonu
\er
where $\e^{\tilde \g} = (\cosh \th - \sin \vth_p)/(\cosh \th + \sin \vth_p)$.

Let $\l^{(1)} = \l^{(2)} = 2$,
$\d (1,0) = z_1 (\L_{r-1} + (-1)^{r/2} \L_r)$
and $\d(0,1) = z_1 (\L_{r-1} - (-1)^{r/2} \L_r)$, then:
\br
  && \d (2,0)/z_1^2=\d (0,2)/z_2^2= (0,0,1, -1, 1, -1,\cdots,(-1)^{r-1},0,0)\cr
  && \d (1,1)= z_1 z_2{1+(-1)^r \over r-1} {\cosh \th \over \cosh \th + 1}
     \L_0 \cr
  && \qquad\quad
    +z_1 z_2\sum_{s=1}^{r-2}\; {2 \cos^2 \vth_s \(1+(-1)^{(r+s+1)}\) \over r-1}
       {\cosh \th - 2 \sin^2 \th_s \over \cosh \th - 2 \sin^2 \vth + 1}\L_s \\
  && \d (2,2) = \d_0^2(1,1)\ u. \nonu
\er
Taking the static limit of the above double--soliton solution, and letting
$y_1 = z1 + z2$ and $y_2 = (-1)^{r/2} ( z_1 - z_2)$, it will  then have
single--soliton form with $\d^{(m)} = \sum_{m_1 + m_2 =m} \d(m_1, m_2)$,
which gives the $\k=2$ solution of \rf{b2}.

We are interested in presenting an N--soliton solution. As in section 3,
it will be built on some $(N-1)$--soliton solutions, and for simplicity,
we assume that all solitons involved are of the form \rf{b1}. Here we
list all the non--trivial coefficients:
\br
  \mv  \d_0 (1,1,\cdots,1)= \d_1 (1,1,\cdots,1) =\prod_{a<b} \d_{ab,0}(1,1),
    \lab{b0111} \\
  \mv \d_r (1,1,\cdots,1)= \d_{r-1} (1,1,\cdots,1) =(-1)^{s_1+s_2+\cdots + s_N}
    \d_0 (1,1,\cdots,1),     \lab{br111} \\
  \mv \d_j (1,1,\cdots,1)=   \sum_{\{\s\}} \displaystyle
    { \cos \Bigl( (2j-1) (\s_1 \vth_{s_1} + \s_2 \vth_{s_2} + \cdots
    + \s_N \vth_{s_N} ) \Bigr) \over \cos \vth_{s_1} \cos \vth_{s_2}
    \cdots \cos \vth_{s_N} } \prod_{a<b} \e^{\g_{ab}^{\s_a \s_b}},
    \lab{b111}\\
  \mv \d_j (\overbrace{2,\cdots,2}^i,1,\ldots,1)= \d_j (2,\ldots,2,0,\ldots,0)
    \d_j (0,\ldots,0,1,\ldots,1) \prod_{a\leq i \atop  b>i} \d_{ab,0}(1,1),
    \lab{b221} \\
  \mv \d_j(2,\ldots,2) = \d_j^{(2)} \d_0^2 (1,\ldots,1). \lab{b222}
\er
In Eqs.\ \rf{b111}, \rf{b221} and \rf{b222}, $j$ runs from 2 to r-2.
We sketch our construction below.

One can easily write down the second--order difference equation for
$\d_j(2,\cdots,2)$ from the vanishing conditions of $V_j(4,\cdots,4)$:
\be
   \d_j^2 (2,\cdots,2) - \d_{j-1} (2,\cdots,2) \d_{j+1} (2,\cdots,2) = 0,
      \quad \mbox { for j = 3, 4, $\cdots$, r-3. } \lab{B5}
\ee
We need two initial conditions to solve \rf{B5}:
\br
   && \d_0(2,\cdots,2)=0, \quad \Longrightarrow \quad
      \d_2 (2,\cdots,2) = - V_0(2,\cdots, 2) = \d_0^2(1,\cdots,1), \cr
   && V_2 (4,\cdots, 4)=0, \quad \Longrightarrow \quad
      \d_3 (2,\cdots,2) = \d_2 (2,\cdots,2) = \d_0^2(1,\cdots,1). \nonu
\er
{}From these two conditions and \rf{B5} we see that \rf{b222} is valid.

One can use the vanishing conditions of $V_j(4,\cdots,4,1,\cdots,1)$ or
$V_j(4,\cdots,4,3,\cdots,3)$  to write down a second--order difference equation
for $\d_j(2,\cdots,2,1,\cdots,1)$, for $3\leq j \leq r-3$.
The two initial conditions are:
\br
   && \d_0(2,\cdots,2,1,\cdots,1)=0, \quad \Longrightarrow \quad
      \d_2 (2,\cdots,2,1\cdots,1) = - V_0(2,\cdots, 2,1,\cdots,1) , \cr
   && V_2 (4,\cdots, 4,1,\cdots,1)=0. \nonu
\er
Factoring out $\d_j (2,\ldots,2,0,\ldots,0) \prod_{a\leq i< \ b}
\d_{ab,0} (1,1)$ from the equations and the conditions, and comparing them to
$V_j(0,\cdots,0,1,\cdots,1)$, we see that \rf{b221} is valid.

One can write down the second--order linear difference equations for
$\d_j (1,\cdots,1)$, for $3 \leq j \leq r-3$.
They are in the  same form as in the $A_r^{(1)}$ case.
Choosing the y's in \rf{a111} to be $y_a^{\s_a}=\e^{-i\s_a \vth_a}$
and comparing  to
\rf{b111},  one sees that \rf{b111}  satisfies the difference
equation. However, this does not prove \rf{b111} uniquely; one still
has to verify that the boudary conditions are indeed satisfied.

%%%%%%%%%%%%%%%%%%%%%%%%%%%%%  Conclusions %%%%%%%%%%%%%%%%%%%%%%%%%%%%%%%%%%
\section{Concluding Remarks}

Hirota's method has been used to construct exact solutions for affine
Toda theories.  We have endeavored systematically to explore this
iterative method and although we do not claim completeness, we have
presented new, explicit multi--soliton solutions for all the cases
discussed, including the detailed procedures for how they were
obtained. The double--soliton solutions given can be proved
explicitly, while general N--soliton solutions are hard to verify
directly.  Scattering solutions of $E_6$, $E_7$ or $E_8$ affine Toda
solitons can also be obtained in this way, due to the lengthiness of
the expressions, they are not included here.  In addition to the
simply--laced loop algebras which we discussed in detail, we have also
given some examples of multi--soliton solutions of non--simply--laced
ones. These can be solved directly using the iterative procedures, or
indirectly, as we have done, by twisting the simply--laced solutions.

For simply-laced affine TFTs, the $\k=2$ or 3 solitons are actually
composites of more fundamental $\k=1$ solitons, and can be obtained by
taking the zero--velocity limit of some scattering solutions, even
though they have single--soliton form.  For non-simply-laced affine
TFTs, the $\k=2$ or~3 solitons are just as fundamental as $\k=1$
solitons.

After this phase of our work was completed, additional recent work by
us \ct{us} cast considerable doubt that many of the soliton solutions
actually have real energy density, contrary to the hopes expressed by
others \ct{hollo1,Evans}. While not negating the interest of the
multi--soliton solutions presented here vis-\` a-vis the knowledge
they give us about the detailed classical spectrum of the ATFTs per
se, nevertheless, these problems about real energy densities may cause
difficulties in considering the the physical applications of some of
the theories.

The single--soliton solutions have a continuous parameter $\xi$, which
is related to the shape and topological charge of the soliton.  For
most choices of $\xi$--value, the energy density is a complex function
of $x$ and $t$. By demanding the energy density of a single--soliton  be
real, then $\xi$ can only take discrete values. However, the reality is
not always preserved in the scattering. All this will be explored
further in a future publication \ct{us}.

After our work was completed, we received the paper of Olive {\it et.\
al}\/  \cite{otu2}, in which a different method, based on Vertex
operators, for finding soliton solutions for TFT's is developed. It
will be interesting to compare explicit results found with their
method to our results.

%%%%%%%%%%%%%%%%%%%%%%%%%%%%%%%%%%%%%%%%%%%%%%%%%%%%%%%%%%%%%%%%%%%%%%%%%

\end{document}